\documentclass[conference]{IEEEtran}
\IEEEoverridecommandlockouts
\usepackage{cite}
\usepackage{amsmath,amssymb,amsfonts}
\usepackage{algorithmic}
\usepackage{graphicx}
\usepackage{textcomp}
\usepackage{xcolor}
\def\BibTeX{{\rm B\kern-.05em{\sc i\kern-.025em b}\kern-.08em
    T\kern-.1667em\lower.7ex\hbox{E}\kern-.125emX}}

\usepackage{hyperref}
\usepackage{changes}

\usepackage{multirow}
\usepackage{array}

\newcolumntype{P}[1]{>{\raggedright \arraybackslash}p{#1}}
\newcolumntype{C}[1]{>{\centering \arraybackslash}p{#1}}

\begin{document}

\title{Sustainability in Telecommunication Networks and Key Value Indicators: a Survey
\thanks{
A preliminary version of this work has been presented at the 2024 IEEE International Symposium on Personal, Indoor, and Mobile Radio Communications (IEEE PIMRC), with a paper entitled ``Building the Foundations of Ethical Networks: Integrating Key Value Indicators for Social, Economic, and Environmental Impact" \cite{Pintor24}
}
}

\author{\IEEEauthorblockN{1\textsuperscript{st} Lucia Pintor}
\IEEEauthorblockA{\textit{UdR Cagliari} \\
\textit{CNIT}\\
Cagliari, Italy \\
ORCID: 0000-0001-7415-6612}
\and
\IEEEauthorblockN{2\textsuperscript{nd} Luigi Atzori}
\IEEEauthorblockA{\textit{DIEE} \\
\textit{University of Cagliari}\\
Cagliari, Italy \\
ORCID: 0000-0003-1350-3574
}
\and
\IEEEauthorblockN{3\textsuperscript{rd} Antonio Iera}
\IEEEauthorblockA{\textit{DIMES} \\
\textit{University of Calabria}\\
Cosenza, Italy \\
ORCID: 0000-0001-5414-8252}
}

\maketitle


\begin{abstract}
Telecommunication technologies are important enablers for both digital and ecological transitions. 
By offering digital alternatives to traditional modes of transportation and communication, they help reduce carbon footprints while improving access to fundamental services. 
Particularly in rural and remote areas, telecommunications facilitate access to education, healthcare, and employment, helping to bridge the digital divide.
Additionally, telecommunications can promote sustainability by supporting renewable energy usage, gender equality, and circular economies. 
However, defining the role of telecommunications in sustainability remains complex due to the historical focus on performance rather than long-term societal goals.
Given the significance of this theme, this paper aims to provide the reader with a deeper look at the concept of sustainability within the telecommunications sector by examining relevant initiatives and projects.
It reviews the major approaches for measuring sustainability and outlines practical approaches for implementing these assessments.
Furthermore, the paper explores the proposed network architectures that incorporate Key Value Indicators and discusses major technologies in this area, such as Network Digital Twins and Intent-Based Networking. 
Through this analysis, the paper aims to contribute to creating sustainable telecommunication networks and broader industries.

\end{abstract}



\section{Introduction}
\label{sec:introction}
Telecommunications play a key role in digital and ecological transitions because they provide (directly and indirectly) an alternative to traditional transportation of goods and people by carrying information electronically. 
Some examples are effective distance meetings that prevent people to move for encountering colleagues, people working remotely without the need to commute to overcrowded cities, digital versions of goods (whenever possible) being transmitted instead of physical ones, online surgeries being performed by doctors without the need for the patients to travel, students attending online courses and performing hands-on activities.   
Decreasing travel leads to lower greenhouse gas emissions, helping to fight climate change. 
Moreover, telecommunications are fundamental for the development of rural areas, reducing the digital divide and creating jobs and economic opportunities.
They also facilitate the establishment of infrastructure for education and culture, such as schools and libraries, and improve access to healthcare.

As telecommunications have become overall present components in all human activities, they can also drive the sustainability revolution from other points of view. 
In particular, they can be deployed and used to create relevant services in a way that favors sustainable deployment and limits approaches that only look for short-term private advantages of restricted groups of stakeholders. 
This aspect may concern using renewable energies, promoting products that endorse environmental preservation and gender equality, supporting a circular economy, and giving everybody the same opportunities. 
However, the role of the telecommunication networks in this respect is not always clear and easy to manage due to the fact that in the past, their development has been driven by only performance indicators that have only sometimes been linked with sustainability. 
Still, in the last years, key initiatives have started to study this topic, and significant projects have been conducted in this respect in terms of standardization, development of tools, and design of concrete use cases that need to be taken properly into account to foster further progress in this direction. 

Given the importance of the theme, this paper aims to provide the reader with a deeper look at the concept of sustainability, describing relevant initiatives and projects in the telecommunications and other sectors. 
Lately, it reviews the major approaches that have been proposed to conduct relevant measurements of sustainability and how to implement them. 
Accordingly, the main contributions of this work are the following:
\begin{itemize}
    \item Different perspectives in sustainability are provided together with the major standardization bodies that have a key role in defining the objectives, the principles, the stakeholders, and the methodologies to achieve the targets. A special focus is given to sustainability in the telecommunications sector \ref{sec:sustainable_overview}).
    
    \item 
    The Key Values (KVs) that are adopted in conducting a business enterprise when societal objectives are taken into account are later depicted. 
    Among the various stakeholders involved in promoting these values, users are the primary drivers, as their demands and preferences can propel businesses toward achieving these key values (Section \ref{sec:key_values}).
    
    \item The Key Value Indicators (KVIs) are then discussed. Even though they define the objectives in terms of measurable achievements, an overall accepted list of KVIs does not exist yet. Recently, many proposals have been proposed from different fora and projects, so further time is needed to achieve a stable and globally accepted view (Section \ref{sec:kvi}).  
    
    \item Three network architectures supporting KVIs are reviewed, taking into account the KVs of the involved stakeholders and verifying that the network is able to achieve the proposed goals. 
    Some key technologies are discussed, such as Network Digital Twin and Intent-Based Networking (Section \ref{sec:architecture}).
\end{itemize}

A discussion of the final issues is then provided at the end of the paper (\ref{sec:future_issues}), with reference to: the certification of the achievement of desired KVIs by the service provider; the usage of the user digital twin to understand the KVIs of interest for the users; managing contrasts among KVIs required by different stakeholders; the translation of KVIs into proper network configurations; assuring the achievement of KVIs in an end-to-end setting.

\section{Sustainability: Concepts and Projects}
\label{sec:sustainable_overview}
\begin{figure}
    \centering
    \includegraphics[width=0.8\linewidth]{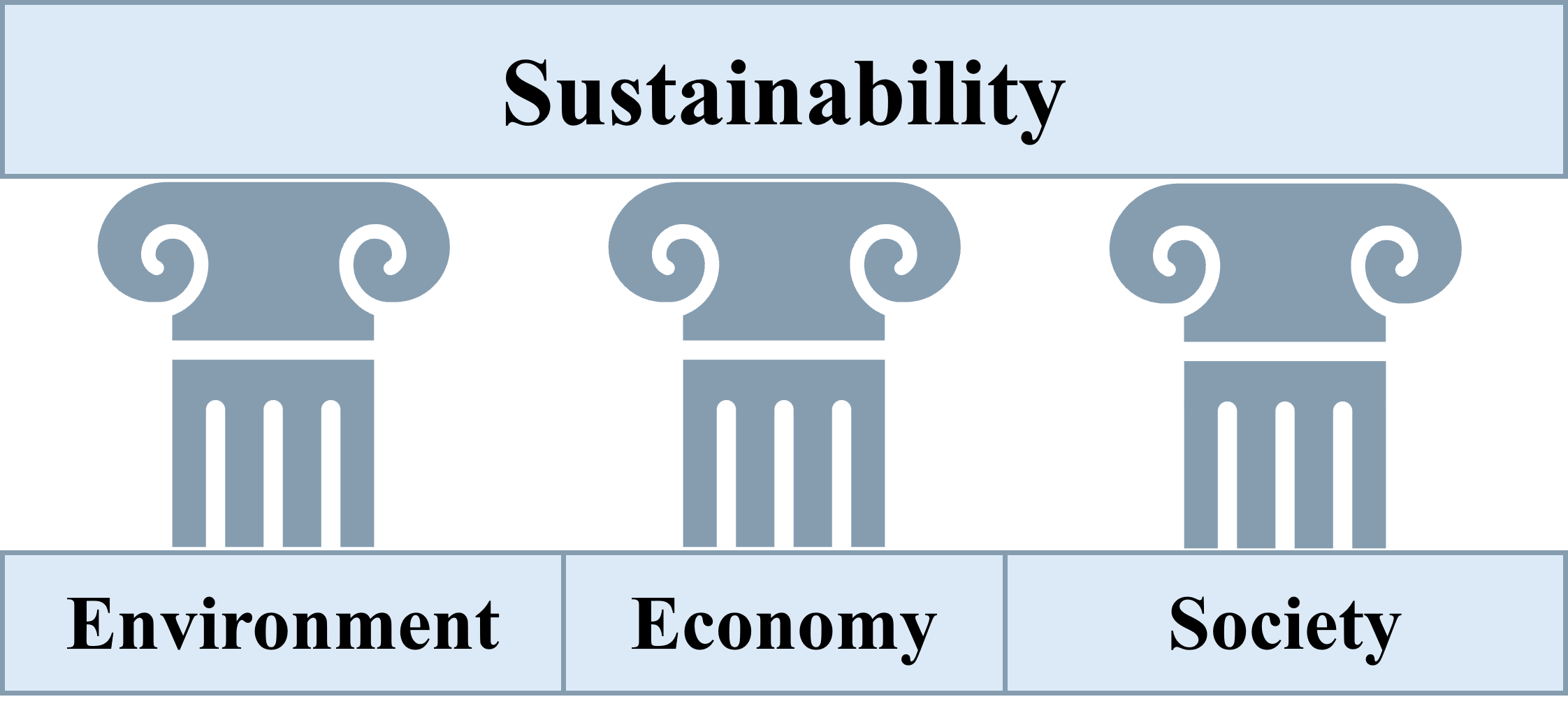}
    \caption{The pillars of sustainability}
    \label{fig:3pillars}
\end{figure}

Sustainability is a development model that aims to satisfy our present needs without compromising the ones of future generations. 
It is a multidisciplinary approach that evaluates the social, environmental, and economic impacts, as shown in Figure \ref{fig:3pillars}.
The well-being of the environment, economies, and people are connected and require global cooperation \cite{Strange08}.
Economic sustainability supports long-term economic growth without negatively impacting social, environmental, or financial systems. 
Social sustainability aims to create healthy and livable communities. 
Environmental sustainability is related to climate stability, waste management, and responsible resource utilization. 
Neglecting even one of these aspects leads to unsustainable choices. 
One example is focusing only on profit, which during the first and second industrial revolutions has proven to be detrimental to the environment and society. 
Sustainable development is a process that changes priorities from the short to the long term.
A balance is expected between current welfare and what would be beneficial in the future to prevent irreparable damage.
Moreover, a single community or country cannot address an issue of this magnitude because it affects the whole world.
This section introduces some of the major initiatives supporting sustainable development and different points of view of vertical markets. It concludes with a focus on the telecommunication field.

\subsection{Major Initiatives about Sustainable Development}
The importance of global sustainable development is attested by the various international agreements defined on this topic over the years.
The Stockholm Declaration (1972) \cite{stockholm1972} was the first major international gathering to address environmental issues on a global scale.
The Brundtland Report, officially titled ``Our Common Future" (1987) \cite{brundtland1987}, forged the term ``sustainable development" and stated that critical global environmental problems were primarily the result of the enormous poverty of the South and non-sustainable patterns of consumption and production in the North. 
Other important treaties adopted subsequently, such as the  Kyoto Protocol (1997) \cite{kyoto1997} and the Paris Agreement (2015) \cite{paris2015}, focus on reducing climate change.
One of the most recent agreements, the UN Agenda 2030 \cite{sdgs2015}, finally considers the three dimensions of sustainable development together through 17 objectives addressing various issues, which are the \textit{Sustainable Development Goals} (SDGs).
Table \ref{tab:sdgs} presents the complete list of SDGs.

Since the adoption of the Agenda 2030, countries have started to incorporate the SDGs into national plans and strategies, private entities have begun to move away from profit-oriented to sustainable-oriented business models. Also, civil society and non-governmental organizations are raising awareness about this topic. 
Initiatives like the European Green Deal \cite{eu_green_deal} (2019), the Inflation Reduction Act of the United States \cite{us_inflation_reduction} (2022), the Australian Net Zero Plan \cite{au_net_zero_plan} (2022), the African Agenda 2063 \cite{africa_agenda2063} (2015) emphasize the global commitment to sustainable development.
The European Union became a reference institution in defining rules and regulations supporting the development of the 2030 Agenda, leading many alliances with other countries like the Africa – Europe Alliance for Sustainable Investment and Jobs (2018), the EU climate cooperation with the Americas (2022), the EU and Japan Green Alliance (2021), the EU and Republic of Korea Green Partnership (2023), and many others.
The SDGs define goals at the level of nations, yet industry and business play a crucial role in their realization.

\begin{table}[h!]
    \centering
    \caption{Sustainable Development Goals of the United Nations}
    \begin{tabular}{|c|l|}
        \hline
        \textbf{SDG ID} & \textbf{Description} \\ \hline
        1             & No Poverty \\ \hline
        2             & Zero Hunger \\ \hline
        3             & Good Health and Well-being \\ \hline
        4             & Quality Education \\ \hline
        5             & Gender Equality \\ \hline
        6             & Clean Water and Sanitation \\ \hline
        7             & Affordable and Clean Energy \\ \hline
        8             & Decent Work and Economic Growth \\ \hline
        9             & Industry, Innovation, and Infrastructure \\ \hline
        10            & Reduced Inequality \\ \hline
        11            & Sustainable Cities and Communities \\ \hline
        12            & Responsible Consumption and Production \\ \hline
        13            & Climate Action \\ \hline
        14            & Life Below Water \\ \hline
        15            & Life on Land \\ \hline
        16            & Peace, Justice, and Strong Institutions \\ \hline
        17            & Partnerships for the Goals \\  \hline
    \end{tabular}
    \label{tab:sdgs}
\end{table}

\subsection{Sustainability in Different Vertical Markets}
Sustainability is a concept that concerns all areas of human activity. 
Among the various industrial sectors, several approaches to implementing sustainable development can be highlighted.
The sectors most significantly impacted by the environmental aspects are the ones related to energy production and management, particularly addressed by SDG 7 (Affordable and Clean Energy), which made significant efforts to reduce greenhouse gas (GHG) emissions and improve energy efficiency. 
However, emissions are not the only focus of this goal: actions to allow people to access affordable and reliable energy services and to increase the renewable energy share are required \cite{EPRS_sdg7_2023}.

Similarly, the European Union has implemented a textiles ecosystem transition pathway within the textile sector, which substantially impacts GHG emissions, climate change, and the use of water and land. 
This transition implies the development of eco-design measures to promote circularity in textile products, the use of recycled materials, the management of hazardous chemicals, and the provision of sustainable textile options for companies and consumers. This program also aims to facilitate access to reuse and repair services \cite{EC_textile2022}.

Another sector that has a huge impact on climate change is transport because it is responsible for approximately a quarter of energy-related global GHG emissions \cite{EESC2023}.
On the other hand, transport is an essential enabler of several SDGs, especially SDG 3 (Good Health and Well-being), 9 (Industry, Innovation, and Infrastructure), and 11 (Sustainable Cities and Communities).
Having an efficient transport infrastructure is fundamental to providing access to job opportunities, goods, and services. 
The initiatives in this regard are mostly related to Europe (e.g.,  \href{https://cinea.ec.europa.eu/news-events/news/horizon-europe-eu1635-million-available-fund-green-smart-and-resilient-transport-and-mobility-2024-05-07_en}{Horizon Europe program 2023-2024}), also given the conformation of its cities with small historic centers not suitable for high traffic flows.
Aspects related to quality education (SDG 4), decent work conditions (SDG 8), and reduced inequalities (SDG 10) are also critical in this context \cite{sdg_hlpf_inputs_2023}. 
These aspects are closely connected because education is foundational for achieving decent work by equipping people with the necessary skills. Quality education and decent work contribute to reducing inequalities by ensuring equitable access to opportunities.

The telecommunication sector is addressing similar problems to support sustainable development: sustainable networks need affordable and clean energy, a circular economy for the production of the equipment, an equal availability of the services should be guaranteed, and no differentiation in the traffic treatment on the basis of the users should be implemented, to cite a few. These issues are discussed in the following subsection.

\subsection{Sustainability in Telecommunication Networks}
The great influence of telecommunications in industry, commerce, and everyday life has led to the realization of specific projects mainly oriented toward the sustainability of telco networks.
The Horizon Europe program plays a crucial role in funding flagship projects, such as  \href{https://adroit6g.eu/}{ADROIT 6G}, \href{https://hexa-x.eu/}{Hexa-X} project and its continuation \href{https://hexa-x-ii.eu/}{Hexa-X-II}, for advancing telecommunications and next-generation network technologies.
Such projects aim to define the enablers and guidelines for 6G architectures that leverage new technologies and programmable Software Defined Networks (SDNs) driven by Artificial Intelligence (AI) to make networks sustainable at all levels (deployment, management, and applications).

An important institution of reference is the \href{https://www.itu.int/en/ITU-D/Environment/Pages/Priority-Areas/Sustainable-Development-Goals.aspx}{International Telecommunication Union} (ITU), which actively promotes especially Goal 7 (Affordable and Clean Energy), Goal 12 (Responsible Production and Consumption), and Goal 13 (Climate action). 
The Groupe Speciale Mobile Association (GSMA) regularly assesses how the mobile industry is advancing to being net zero by 2050. 
Its last report \cite{gsma2022} documents that even though network traffic increased in the last years, operators were able to reduce their carbon footprint emissions by improving efficiency and by using renewable sources. 
Moreover, based on mobile industry analysis, the main source of emissions for operators is from the industry’s supply chain.
Sustainable network equipment should be modular, interoperable, and designed for a long life cycle. 
Virtualization is the key to centralizing network intelligence and control at the software layer to standardize the underlying hardware.
Moreover, virtualization will allow the implementation of network updates without physically changing the legacy hardware. 

Other important communities of this sector, such as the 6G Smart Networks and Services Industry Association (6G SNS-IA) \cite{6GSNS-IA} and the European Telecommunications Network Operators' Association (ETNO) \cite{etno24}, investigate the adoption of social dialogue to new digital work environments, emphasizing workers' well-being and engagement.

Proactive social dialogue builds trust between workers and companies, making companies more attractive to employees. 
This practice creates a sense of belonging and competitiveness, fostering a beneficial cycle where highly skilled workers align with the company's goals and are motivated to be hired and remain with the company.

\subsection{Discussion about Sustainable Networks}
The concept of sustainable networks encompasses the design, deployment, and management of network architectures that are economically convenient, socially responsible, and environmentally friendly. 
Energy management and resource optimization are crucial for economic and environmental aspects because they allow for minimizing energy consumption and emissions without compromising performance.
Moreover, renewable energy sources and the circular economy drastically reduce the environmental impact.
Despite that, the cost of innovation should provide long-term economic benefits to be cost-effective.
The design of sustainable networks must ensure scalable and dynamic adaptation to market demands and technological advancements.

Additionally, social aspects should not be neglected.
Ensuring equitable access to high-quality network services mitigates the digital divide, reducing differences among communities living in remote areas or with special needs.
Other ethical concerns are the preservation of privacy, transparency and fairness of data processing, and education.
The main challenges to the development and implementation of  \textit{Ethical Network}, i.e., telecommunication networks designed to reflect the ethical principles of sustainability, are due to the need to standardize metrics and protocols to measure the achievement of SDGs.
In this direction, engaging with stakeholders in academia, industry, and governments is mandatory to create a supportive ecosystem that produces value.

\section{Key Value analysis}
\label{sec:key_values}

In business processes, ``value" refers to the benefits or outcomes that a company provides to its customers, stakeholders, or itself through its operations and activities. 
It encompasses the creation, delivery, and capture of products or services that meet the needs and expectations of customers while also contributing to the company’s overall objectives, such as profitability, growth, and sustainability.

\begin{figure}
    \centering
    \includegraphics[trim={9.5cm 0 1.2cm 0},clip, width=1\linewidth]{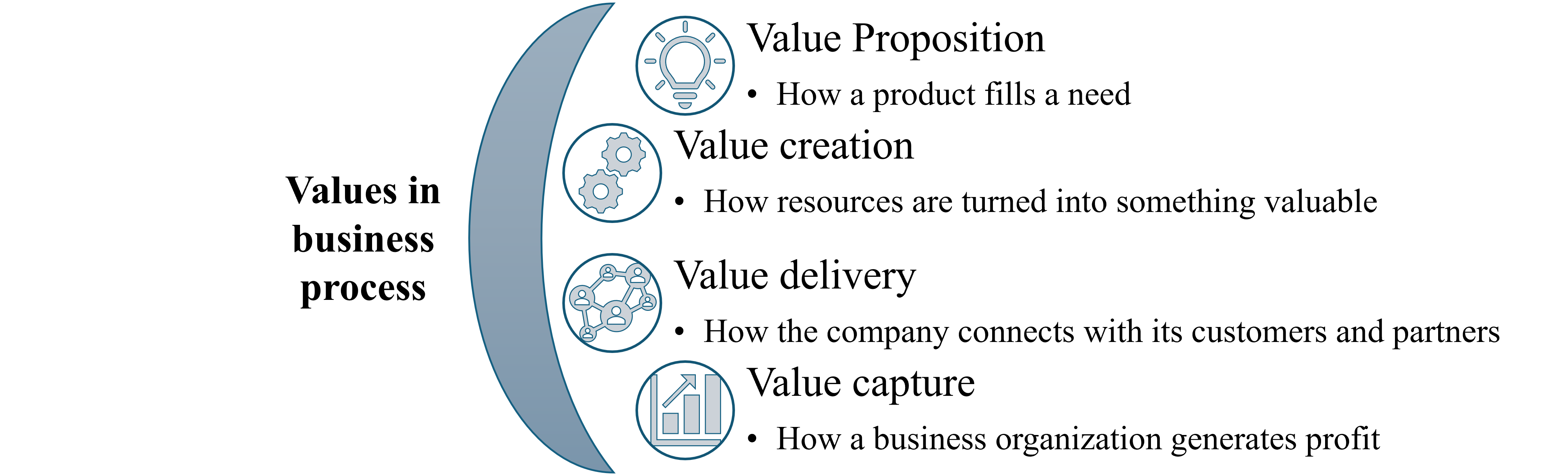}
    \caption{Values in business process}
    \label{fig:business_values}
\end{figure}

The concept of values is also strictly linked with that of the business model, which focuses on four dimensions, as shown in Figure \ref{fig:business_values}: value proposition, value creation, value delivery, and value capture \cite{osterwalder2004business, RestartWP}. 
Value proposition defines how a product fills a need by demonstrating its benefits and comparing it to similar products on the market.
Value creation involves all the activities, resources, and partnerships essential for developing products or services that form the value proposition of a company, targeting various customer segments. 
Value delivery focuses on how the company is structured to connect with its customers and partners, encompassing aspects such as customer relationships, sales, and distribution channels. 
Finally, value capture describes how a business organization retains value and generates profit, including revenue generation, profit-sharing with partners, and managing investment, financing, and cost structures.

In business processes, several ``key values" are typically considered to ensure that the operations align with the goals of a company and provide meaningful benefits to customers and stakeholders. 
Here are some of the most common ones: customer value, which is linked with quality and customer satisfaction; efficiency, in terms of process optimization and resource management; cost-effectiveness, which focuses on cost reduction; sustainability, which focuses on environmental responsibility and long-term viability; profitability; compliance and ethics, which relates to regulatory compliance and ethical standards; scalability; risk management; innovation and continuous improvement adaptability; employee satisfaction and engagement; partnerships and collaboration value chain integration. 
These values guide the design, implementation, and management of business processes, ensuring that they contribute to the overall success and sustainability of the organization \cite{jeston2014business}.


In this context, telco operators are increasingly focusing on sustainable strategies to attract investment primarily but also to increase their appeal to customers and employees \cite{Yrjola2023}.
In the rest of this section, we discuss how Key Values (KVs) \cite{6GSNS-IA} are becoming fundamental principles leading the decision-making processes of organizations or communities through the definition of double materiality, the role of the end-users in the sustainable transition, and how to advertise the efforts of implementing sustainable development practices.

\subsection{Accelerating the Transition to a Sustainable Development through Double Materiality}
Supporting companies that are committed to these goals is not only a matter of corporate social responsibility for market investors but also a strategic investment decision. 
Companies that align with sustainability initiatives are likely to be better prepared for the future regulatory environment and more attractive to consumers who value sustainability.
Putting together the financial interests of the companies and investors and the sustainability long-term perspective leads to the definition of double materiality \cite{Delgado23}.
“Materiality” refers to the information companies must provide investors with.
Having double materiality indicates that companies must report both the economic performance of their business (“financial materiality”) and how their activities impact society and the environment (“environmental and social materiality”) \cite{euclimate2019}.
Even though investors are usually more interested in financial materiality, they are raising their interest in the environmental and social impacts of companies where they invest.
By directing capital towards businesses that support the SDGs, investors can play a crucial role in accelerating the transition to a sustainable future.
Investors can look for firms that demonstrate clear commitments to \textit{Environmental, Social, and Governance} (ESG) criteria.
Rating agencies like Morningstar and S\&P Global evaluate ESG indices that measure all the efforts related to responsible investment that pursue the typical objectives of financial management by considering sustainable aspects.
However, there is no standard to measure and benchmark the general propensity of companies to adopt sustainable development practices.
Furthermore, one must consider that different sectors impact differently on sustainability.

\subsection{The Role of the End-Users}
\label{subsec:users}
Two main levers, other than the investors, can bring the companies to adopt key values criteria in their activities. 
The first one is linked with the customers, who may be more interested in consuming services from companies that reflect their own values once the services satisfy their needs. 
The second one is related to regulations that, at different levels, stimulate companies to adopt the right values in their activities. 
Laws can be a sharp constraint to operate in a given sector or can be an economic incentive provided only once value-related goals have been achieved. 
The two levers can operate in parallel and are expected to be introduced with different timing. 
The customers may be faster than the regulatory process in asking the companies to adapt their operativity. 
This fact is also connected with the mass trend phenomena. 

As to the first lever, the companies need to know which are the values the customers are most interested in. 
In this direction, they may use various methods to gather customer opinions, which help them understand customer satisfaction with the current adopted values, preferences, and areas for improvement. 
Companies often use surveys to ask customers about their experience, satisfaction, and advice. 
This information can also be collected in conjunction with the purchase of a service so that the retailers may ask customers to answer short questionnaires with reference to the preferred values and whether those provided by the companies are those of interest. 
Another interesting approach is to perform social media monitoring so that tools are used to monitor social media platforms for mentions of topics related to the key values and whether they are associated with their brand, products, or services. 
This analysis helps gauge public sentiment and identify trends or issues. 
The evaluation of customer sentiment could also be performed through direct interaction so that businesses engage with customers directly on social media, responding to comments, messages, and reviews to gather insights and address concerns. 
Another important approach is the activation of specific focus groups, where a small number of customers are invited to discuss their opinions in detail. This qualitative method provides in-depth insights. Many other approaches could be used, such as customer interviews, customer support interactions, net promoter score surveys, behavioral analytics, community forums and user groups, competitor analysis, and crowdsourcing ideas.

\subsection{Advertise the Adopted Key Values}
The companies then need to communicate the key values effectively to customers using a combination of different strategies and channels, such as \cite{berger2013contagious, miller2017building}: mission statement and vision displayed on website and other communication channels; adequate product and service design especially showing proper certifications and labels; proper branding and messaging; content marketing (e.g., publish content that discusses the company’s values and how they guide business decisions, create videos that tell the story of the company’s values, run social media campaigns that highlight the company’s values through posts, stories, and customer testimonials).

A pivotal challenge in this respect is to properly communicate to the current and future customers in an effective way the key values chosen by the company and how these are achieved in a transparent, trustworthy, and understandable way.
In this perspective, it becomes important to have standardized methodologies available both to measure the Key Values inspired by sustainability principles and to map specific Key Value Indicators (KVIs) that describe them on performance metrics or Key Performance Indicators (KPIs) for sustainable telecommunications systems.
In the following Sections, the main activities, projects, and studies conducted and currently underway relating to KVIs for future Ethical networks are presented and discussed.

\section{Key Value Indicators: Converting KV into Metrics for Sustainable Telecommunications}
\label{sec:kvi}

Suppose we start from the concept that ``you can only effectively manage what you are able to measure." In that case, we realize the importance of measurement and reporting methodologies for metrics specifically linked to sustainability. 
Many organizations and standardization bodies have faced the problem of providing indications on how to measure sustainability through appropriate KVIs and KPIs. Some of these initiatives are described below.

\subsection{General Frameworks for Measuring Sustainability}

The Global Reporting Initiative (GRI) Sustainability Reporting Standards are a well-known example of standards that provide widely used guidelines for reporting the sustainability performance of companies and organizations of any size, sector, and country in the world.
An important aspect to highlight is that these standards do not define specific KVIs for any industry. 
Instead, as stated in the GRI documents, they represent good practices for public reporting on a range of economic, environmental, and social impacts   
\cite{GRI}.

Specifically, the GRI framework provides a structure for companies to report on their sustainability impacts, including standards from three categories:
\begin{itemize}
    \item Universal Standards that apply to all types of organizations and establish the foundation for reporting but aren't KVIs themselves.
    \item Sector Standards which are based on specific areas. While there is not one for telecom yet, relevant aspects from other sectors might be adaptable (e.g., Oil \& Gas Standards for Energy Consumption).
    \item Specific Standards contain information for the organization regarding the reporting of impacts relating to specific times.
\end{itemize}

The GRI framework provides guidelines for reporting, so although it is reasonable to have some common global indicators (such as emission rates), different industries can contribute to sustainable development in distinct ways by defining specific indicators in line with their main activities and impacts \cite{dvfa2010}.

The Sustainability Accounting Standards Board (SASB) \cite{SASB} provides an additional set of metrics designed to help companies identify, measure, and report their most significant sustainability risks and opportunities.
Even though this framework does not have industry-specific standards for telecommunications companies (as it was the case also for the previous one), it provides a set of sustainability metrics that can be applied to a variety of industries, including telecommunications.
SASB standards include a set of disclosure topics, which vary from industry to industry and describe specific sustainability-related risks or opportunities associated with the activities conducted by entities within a particular industry.


The Task Force on Climate-related Financial Disclosures (TCFD) \cite{TCFD} was a global initiative that aimed to improve the reporting of climate-related financial risks and opportunities by companies. 
TCFD recommendations provide a framework for companies to disclose information that will help investors, lenders, insurers, and other stakeholders assess and price climate-related risks and opportunities more effectively.
The recommendations are structured around four thematic areas that represent core elements of how organizations operate: governance, strategy, risk management, and metrics and targets (the latter in order to track the company's performance on climate-related issues).
TCFD also did not provide a specific set of KVIs for measuring climate-related risks and opportunities in telecommunications systems. 
Currently, the TCFD has fulfilled its mandate, has been dissolved, and has been replaced by the IFRS Foundation \cite{IFRS}, which will be responsible for monitoring the progress of corporate communications related to the climate.

Those illustrated are only the most relevant among the several common ESG reporting initiatives currently underway, some more general and others perhaps more focused on the climate issue. 
Among them, it is worth mentioning the International Integrated Reporting Council (IIRC), Carbon Disclosure Project (CDP), Climate Disclosure Standards Board (CDSB), etc., which are just a few examples of those reported and illustrated in detail in the interesting study by Nordea \cite{Nordea}.  

\subsection{Standards and Guidelines on Measuring KVIs for Telecommunications}


Various organizations are advancing sustainability in the telecommunications sector through standardized frameworks and metrics. 

The International Telecommunication Union (ITU) develops methods for measuring sustainability, including KVI frameworks that assess broader environmental and social impacts beyond traditional KPIs. For instance, ITU-T Study Group 5 focuses on reducing greenhouse gas emissions in digital technologies \cite{ITUSG5}.

The European Telecommunications Standards Institute (ETSI) is committed to a sustainable future and promotes the objectives of the EU industrial strategy to become greener, digital, and resilient. 
To this end, it has established a series of Technical Committees (TCs) that deal with sustainability issues. Among these, the Access, Terminals, Transmission, and Multiplexing (ATTM) committees focus on compliance with the environmental requirements of networks, as well as broadband transmissions.
The ATTM technical committee is also developing global KPIs to enable ICT users to monitor their eco-efficiency and energy management and to monitor the sustainability of broadband solutions \cite{ATTM}.

GSMA, through its Open Gateway Initiative \cite{OGI}, aims to develop models for KVIs suitable for Terrestrial and Non-Terrestrial Network (TN/ NTN) implementations and operations.
Although the GSMA has been performing annual monitoring of the sector's performance with respect to the United Nations SDGs for several years, it has not yet developed sufficiently detailed KVIs for the required measurements.

The Next Generation Mobile Networks Alliance (NGMN) contributes to the topic with a series of white papers on ``Green Future Networks" \cite{NGMN} in which KPIs and respective target values for the evaluation of the green network are also discussed. 
With a view to deploying 6G networks, the approach followed is to guarantee an acceptable user experience while respecting principles of sustainability. 

Similarly, the Next G Alliance (NGA) with its Working Group called Green G, promotes, through their withe-papers (“Green G: The Path Toward Sustainable 6G” \cite{NGA1} and “6G Sustainability KPI Assessment: Introduction and Gap Analysis” \cite{NGA2} ), the environmental sustainability in the development of future wireless technology and the creation of a sustainable 6G ecosystem. The latter are considered enablers of the reduction of greenhouse gas emissions, land and water use, also for other industrial sectors from a circular economy perspective.

\href{https://www.3gpp.org/}{3GPP's} efforts on sustainability and energy efficiency are undertaken by various groups within the organization. For example, Working Group 5 (SA5) of the Technical Specification Group Service and System Aspects, in Release 17, pursues energy efficiency and energy saving of mobile networks, dealing with both aspects related to the Radio Access Network (RAN) and the entire 5G system. 
KPIs in terms of Energy Efficiency are defined for each segment of the 5G network. At the same time, TSG Radio Access Network groups, in the ongoing Release 18, have studied the energy consumption model of the network to identify methods of energy saving for the mobile network and reduce its energy consumption. However, version 19 of the TSG SA Working Group 1 (SA1) aims to introduce the concept of ``energy efficiency as a service". 

Finally, we mention the interesting work conducted by the European Union European Joint Research Center together with DG CNECT in the context of the definition of Key Values described in the document \cite{JRC136475}.
The goal was to develop common indicators for measuring the environmental footprint of Electronic Communications Networks (ECNs) for the provision of communications services. 
Attention is placed on the sustainability of fixed and wireless telecommunications networks in terms of environment, climate, energy consumption, and energy efficiency. It has led to the definition and analysis of 19 sustainability indicators related to the whole life-cycle of networks and equipment and not only the operational phase.

Many telecommunications companies around the world have published sustainability reports in recent years in accordance with the standards we have listed. 
Reporting was done with the aim of demonstrating their commitment to ESG principles and their efforts to minimize environmental impact and promote sustainable practices.
The specific content and format of these reports vary depending on the size, operations, and priorities of the company but typically cover topics such as energy consumption, greenhouse gas emissions, waste management, water use, data privacy of customers, and employee well-being.

\subsection{Academic Research on KVIs for Sustainable Telecommunications}

Even in the academic field, the attention to KVIs is very high, and new interesting works aim to specifically describe the importance of KVIs for the assessment of sustainability for future telecommunications systems, as well as to propose some examples in specific use cases. A non-exhaustive list of some of these works is given below.

\cite{Wikstrom24} presents a study on KVIs as a method for analyzing results relating to values deriving from ICT developments. The authors propose a framework for KVIs composed of five phases, starting from the identification of the values linked to the use case up to the evaluation of the results in terms of value.
The KVI framework is presented as a tool to be used predominantly in research activities to address social challenges in the technological design and development phases and to identify and estimate valuable outcomes resulting from the use of technology.

Instead, the authors in \cite{Osman24} provide a fascinating analysis of the relationships between technology, business, and value by discussing the concept of Value Driven Business Models. They do not define precise KVIs but provide an initial high-level classification by identifying three types: economic, environmental, and social.
In the framework of the research and development activities of the 6G-PATH project, the authors of \cite{Choch} discuss and analyze in detail what they consider to be the ten most relevant use cases. For each of them, they propose sets of KVIs (other than KPIs). 
Their goal is to understand more deeply how many, and which will be the main requirements for the future 6G network.

Similarly, the authors of \cite{wyme} start from a careful analysis of the needs of different stakeholders along the whole value chain to describe and structure the KVIs of a 6G system in the context of ISAC (i.e., communication, positioning, and sensing). 
Synergies and conflicts, as well as methods to quantify such KVIs by transforming them into KPIs, are addressed in the paper.

Differently, the authors of \cite{10152513} discuss the KPIs for this reference domain and map them to the KVIs and the SDGs formulated by the UN, focusing on use cases and applications in the Satellite-IoT field.


\section{Formalizing KVIs to fit a Network Architecture}
\label{sec:architecture}
In this section, we examine the different types of KVIs that could be used in an ethical-oriented network management procedure and the way these could be described. We then review the ones that have been defined in previous studies and projects and the functionalities and technologies that have been proposed to be implemented in future networks for their management.

\subsection{Types of KVI}
\label{sect:KVI_type}
KVIs are metrics that measure the key values defined by an organization, product, or service. 
They focus on a holistic impact and long-term value generation.
They are part of the group of Key Indicators (KIs), which also includes Key Performance Indicators (KPIs), which measure performance in achieving specific objectives, and Key Quality Indicators (KQIs), which assess the quality of products \cite{Atzori23}.
Differently from KPIs and KQIs, two types of  KVIs can be identified:

\subsubsection{Sensor-Based KVIs}
These indicators can be measured and monitored in real time using sensors, Internet of Things (IoT) devices, or other automated systems.
They provide immediate data on the performance and impact of specific activities or processes.
Some of these indicators are similar to KPIs so that they can be treated similarly by an Information Technology (IT) system: a sensor measures the current value as frequently as the metric variates, transmits the information to the gateway, which then makes it available externally via an Application Programming Interface (API).
Examples of sensor-based KVIs are related to environmental monitoring (e.g., pollution emission metrics) and operational efficiency (e.g., resource consumption, waste production, etc.).

\subsubsection{Certified KVIs}
These indicators rely on formal evaluation by certifying bodies to ensure compliance with standards and regulations. 
They often involve a formal assessment or audit process such as surveys.
For this reason, using external certifying bodies ensures greater transparency and reliability, avoiding bias and conflict of interest that occur when evaluated by the same entity that controls the system. 
Certifying bodies should also provide and maintain APIs to retrieve these indices every time they are needed.
While sensor-based KVIs provide immediate feedback, certified KVIs are typically assessed periodically, reflecting longer-term compliance and strategic alignment.

\subsection{Structuring KVIs}
Key Indicators must be identifiable and shared across the network, requiring a centralized database for definitions and distributed databases for measurements. 
Unique identifiers link these databases. 
The centralized database includes the description of each indicator in terms of meaning, encoding, and target values that can be expressed as a single value or a range \cite{Dominguez19}, as shown in Table \ref{tab:ki_char}.
Indicators may have an optimal range of values, so a maximum and minimum should be given. In case there is no lower or upper limit, the field can be left blank. Otherwise, if the target is a specific value, both fields must have the same value.
Indicators might belong to multiple categories, so the best solution for assigning categories might be to define a dictionary of categories.
Each indicator can belong to several categories, and each category can be assigned to more than one indicator.
Moreover, some indicators may only apply under specific conditions, necessitating alternatives for special network states like recovery. 
The field labeled ``condition" stores one or more conditions that play the role of filter, such as constraints or requirements.
It's also important to document relationships between low-level and high-level indicators using a parent-child structure commonly used in processes. 
A parent indicator is a high-level indicator that is based on the measurement of multiple independent low-level indicators.
There is only a children column in the database to simplify the structure.
Knowing the method used to measure the indicator, such as device, API, or formula, can help define it better. 
The source of an indicator indicates the entities and data that are required to calculate it.
While a structured database is useful for containing general information about the indicators, the distributed database could be unstructured and contain current values, time references, and sources.

\begin{table*}[h!]
    \centering
    \caption{Features of a Key Indicator (KI)}
    \begin{tabular}{|l|l|l|}
        \hline
         \textbf{Feature} & \textbf{Description} & \textbf{Values} \\ \hline
         
         ID & A unique identifier of the KI in the centralized shared database. & integer number \\ \hline
         
         Description & A clear and precise description of the KI. & 
         text  \\ \hline
         
         Encoding & The data type of the KI. & string \\ \hline
         
         Type & The type of KI. & KPI, KQI, or KVI \\ \hline
         
         Unit Measure & The basic unit of measure in which the KI is expressed. & string \\ \hline
         
         Timing Context & The type of sampling. & real-time, on-demand, or periodically  \\ \hline 
         
         Frequency & The periodicity of calculation of the KI. & number (or empty) \\ \hline
         
         Min & The minimal target value. & number \\ \hline
         Max & The maximal target value. & number \\ \hline
         
         Condition & 
         The required state of the system. & 
         e.g., ordinary, recovery, etc.   \\ \hline
         
         Categories & Keywords that categorize the KI. & list of tags   \\ \hline
         
         Children & List of IDs of the low-level KIs that compose the actual KI. & list of integer numbers (or empty) \\ \hline
     
         Source & Sensor or API for the measurement or formula to calculate it. & text \\ \hline
    \end{tabular}
    \label{tab:ki_char}
\end{table*}

\subsection{KVIs from Major Research Projects}
\label{sec:KVI_projects}
Several projects have defined Use Cases and their respective KVIs. 
In this subsection, we present a summary of the KVIs presented by the projects ADROIT6G \cite{adroit6g} and Hexa-X \cite{hexax-D1.4}. 
Furthermore, we provide an overview of the KVIs-related activities of the Smart Networks and Services 6G Infrastructure Association (SNS 6G-IA), which incorporates contributions from European stakeholders to provide a European vision on 6G use cases. 
Finally, as some European governments, through their national research programs, are pushing towards sustainable 6G solutions, we illustrate examples of ideas emerging in this context referring to the COHERENT project \cite{Becattini24}, specifically focused on the creation of Network Digital Twins solutions to support future network management consistent with sustainability goals.

\subsubsection{KVIs defined by ADROIT6G}
The ADROIT6G project identifies 15 KVIs listed in Table \ref{tab:kvi_adroit6g} and maps them according to the type of traffic of their Use Cases: i) extreme eMBB (Enhanced Mobile Broadband), ii) extreme mMTC (Massive Machine Type Communications) and NTN (Non-Terrestrial Networks), and iii)  extreme URLLC (Ultra-Reliable Low-Latency Communications) and mMTC.
KVIs are classified into three categories: innovation, democracy, and ecosystem.
The ``innovation" category assesses aspects such as safety, security, regulatory compliance, energy efficiency, and ethical responsibility. 
The ``democracy" measures how well a network promotes inclusivity, fairness, user rights, and trustworthiness. 
The ``ecosystem" category includes indicators like sustainability (herein, with this, they refer only to the environmental impact), business value, economic growth, open collaboration, and new value chains. 
This category ensures that the network contributes positively to the environment and drives economic development while promoting partnerships and the creation of new value chains within the ecosystem.


\begin{table*}[h!]
\centering
\caption{KVIs from the ADROIT6G project \cite{adroit6g}}
\begin{tabular}{|l|l|P{15 cm}|}
\hline
\textbf{ID} & \textbf{Category} & \textbf{Description} \\ \hline

I1 & Innovation & \textbf{Safety} measures the consequences of unavailability or degradation of communication services through surveys. \\ \hline

I2 & Innovation &  \textbf{Security} considers the vulnerabilities and security concerns through questionnaires and analysis of the security issues logs. \\ \hline

I3 & Innovation &  \textbf{Regulation} evaluates adherence to regulatory standards and frameworks in the development and deployment of innovative technologies. \\ \hline

I4 & Innovation &  \textbf{Responsibility} tracks ethical responsibility and accountability in network innovation, ensuring that developments serve societal needs. \\ \hline

I5 & Innovation &  \textbf{Energy Efficiency} measures how innovative solutions contribute to reducing energy consumption and increasing efficiency in network operations through the monitoring of energy consumption per bit. \\ \hline

D1 & Democracy &  \textbf{Privacy} monitors the preservation of user data and privacy rights by identifying potential privacy issues. \\ \hline

D2 & Democracy & \textbf{Fairness} assesses how the network performs decisions free from discrimination and bias. \\ \hline

D3 & Democracy & \textbf{Digital Inclusion} measures the digital divide between people who have access and those who do not. \\ \hline

D4 & Democracy & \textbf{Trustworthiness} evaluates through interviews and surveys the trust users have in the network’s transparency, ethical operations, and reliability. \\ \hline

E1 & Ecosystem & \textbf{Sustainability} tracks the network’s environmental impact in terms of CO\textsubscript{2} emissions per year. \\ \hline

E2 & Ecosystem & \textbf{Business Value} measures the economic benefits generated by the network, including contributions to business innovation and growth. \\ \hline

E3 & Ecosystem & \textbf{Economic Growth} estimates the broader economic impact of the network on national or global markets, considering the investment in education and digital transformation. \\ \hline

E4 & Ecosystem & \textbf{Open Collaboration} measures the engagement in partnerships and collaborations with stakeholders. \\ \hline

E5 & Ecosystem & \textbf{New Value Chain} evaluates the creation of new business models, value chains, and economic opportunities stemming from network innovations. \\ \hline
\end{tabular}
\label{tab:kvi_adroit6g}
\end{table*}

\subsubsection{KVIs defined by Hexa-X}
Hexa-X projects follow an approach that links KVIs, also called proxies, directly to four Key Values: Sustainability, Inclusion, Flexibility, and Trustworthiness. 
Sustainability in Hexa-X refers to minimizing the environmental impact of communication networks by improving energy efficiency, reducing carbon emissions, and optimizing the use of resources. 
Inclusion (or digital inclusion) emphasizes equitable access to digital technologies and services, ensuring that everyone, regardless of socioeconomic background, geographical location, or physical ability, can participate in and benefit from 6G innovations. 
Flexibility involves scalable and configurable network architectures that can efficiently accommodate varying demands, from ultra-reliable low-latency communication (URLLC) to massive machine-type communication (mMTC), ensuring that the network can evolve dynamically.
Trustworthiness aims to build user confidence in the system by safeguarding data integrity, protecting against cyber threats, and ensuring that technologies operate in a fair and accountable manner.
Table \ref{tab:kvi_hexa} shows the KVIs and the enablers defined by the Hexa-X project.
Controversially, Hexa-X-II \cite{hexaxII-D6.3} considers KVIs the Key Values of Hexa-X, Sustainability, Inclusion, Flexibility, and Trustworthiness. 

\begin{table*}[h!]
\centering
\caption{KVIs from the Hexa-X project \cite{hexax-D1.4}}
\begin{tabular}{|l|l|P{9 cm}|}
\hline
\textbf{ID} & \textbf{Key Value} & \textbf{Enablers} \\ \hline

Energy consumption during operation & Sustainability &  Extending sleep modes and joint optimization of computation and communication \\ \hline

Energy consumption at zero load & Sustainability & Advanced sleep modes, low-energy devices, and complexity reduction in signaling \\ \hline

Signaling overhead and complexity & Sustainability & AI-based orchestration, functional modularization, and reduced signaling \\ \hline

Energy consumption per bit  & Sustainability & Control-Computation-Communication-Co-Design \\ \hline

Increase in addressable workforce   & Inclusion & Novel Human Machine Interfaces (HMIs) and DTs to support remote work \\ \hline

Perceived quality of work (and related QoE) & Inclusion & Novel Human Machine Interfaces (HMIs) and DTs to support remote work \\ \hline

Reduction in wait time & Inclusion &  N/A \\ \hline

Ease of use & Inclusion & N/A \\ \hline

Percentage of population reached & Inclusion &  NTN for inclusive E-Health and institutional coverage \\ \hline

Percentage of target area covered & Inclusion & NTN and zero-energy devices for earth monitoring \\ \hline

Spatial and/or temporal Resolution & Inclusion &  NTN and zero-energy devices for earth monitoring \\ \hline

Convergence time & Flexibility & Flexible resource allocation \\ \hline

Detection time & Flexibility &  Handling unexpected situations, and error detection  \\ \hline

Re-configuration overhead & Flexibility & Flexible networks\\ \hline

Re-configuration capability/related to scalability & Flexibility & Flexible resource management and scalability \\ \hline

Re-use/sharing of spectrum  & Flexibility &  Flexible spectrum management \\ \hline
   
AI privacy & Trustworthiness & Privacy-preserving clustering and differentially private Federated Learning (FL) \\ \hline

AI agent availability and reliability & Trustworthiness & Prediction of mobility, workload movement optimization, and prediction of impairments in connectivity  \\ \hline
\end{tabular}
\label{tab:kvi_hexa}
\end{table*}

\subsubsection{KVIs defined by 6G SNS IA}
The approach defined by the SNS 6G-IA White Paper \cite{6GSNS-IA} identifies Key Values related to Use Cases and defines specific KVIs for each KV, as shown in Table \ref{tab:kvi_sns}.
The UC1 (Emergency response \& warning systems) aims to support the Public Protection and Disaster Relief (PPDR), which operates in three stages: incident response (highest priority communication), prevention (limiting impact), and recovery (where communication delays are tolerable).
The UC2 (Smart city with urban mobility) has the purpose of improving the efficiency and accessibility of traditional services, such as urban transport, public safety, healthcare, water supply, and waste management. 6G technology is an enabler to enhance smart cities to provide accessibility of sensor data, combined with AI and data analytics, allowing for predictive city behavior and resource management.
The UC3 (Personal Health Monitoring \& Actuation Everywhere) addresses the challenge of enhancing access to professional healthcare for everybody, which translates into methods for preventive monitoring health and telemedicine.
The UC4 (Living and Working Everywhere) aims to reduce the digital divide by enabling remote education and work, reducing the need for relocation, and providing broader access to cultural products and activities.
The UC5 (Assistance from twinned cobot) is dedicated to collaborative robots (or cobots) that are expected to assist in diverse sectors, including manufacturing, elderly care, and manual labor.
The UC6 (Sustainable food production) focuses on reducing the impact of climate change by finding sensor and digital twin-based solutions to monitor environmental indicators and support crops and livestock farming.
The six Use Cases share several Key Values that, in this case, focus primarily on the social aspects of sustainability.

\begin{table*}[h!]
    \centering
    \caption{Key Values from 6G SNS IA \cite{6GSNS-IA}}
    \begin{tabular}{|P{3 cm}|P{7.5 cm}|P{6.5 cm}|}
    \hline
        \textbf{Key Value} & \textbf{Key Value Indicators} & \textbf{Use Cases}  \\ \hline
        Cultural connection & Access to cultural products and cultural events & Living and working everywhere (UC4) \\ \hline
        Democracy & Access to  and active participation in administrative and political functions  & Living and working everywhere (UC4)  \\ \hline
        Digital inclusion & Access to the internet in communities and areas & Living and working everywhere (UC4)  \\ \hline
        Economical sustainability and innovation & Cost-efficiency of rural living and the range of activities that can be performed & Living and working everywhere (UC4)  \\ \hline
        Environmental sustainability & Area of protected natural habitats and climate preserves, environmental footprint of urban transport and agriculture activities & Emergency response \& warning systems (UC1), Smart city with urban mobility (UC2), and Sustainable food production (UC6)  \\ \hline
        Knowledge & Access to education & Living and working everywhere (UC4) \\ \hline
        Personal freedom & Degree of influence over your daily activities and  degree of personal mobility & Assistance from twinned cobots (UC5)  \\ \hline
        Personal health and protection from harm & Operational efficiency,  access to autonomous health monitoring services,  injuries in urban traffic and labor-intensive activities & Emergency response \& warning systems (UC1), Smart city with urban mobility (UC2), Personal health monitoring \& actuation everywhere (UC3), Assistance from twinned cobots (UC5)  \\ \hline
        Privacy and confidentiality & Reported user control of medical data for storage/transmission/processing & Personal health monitoring \& actuation everywhere (UC3)  \\ \hline
        Simplified life & Improved access and ease of use of public transport, along with time savings in agricultural activities & Smart city with urban mobility (UC2) and Sustainable food production (UC6)  \\ \hline
        Societal sustainability & Emergency response times and operational efficiency in remote areas, cost savings in healthcare, reduced travel times, access to job markets and life opportunities in rural areas,  cost-efficiency in labor-intensive operations, and  agricultural productivity and reliability of food production & Emergency response \& warning systems (UC1), Personal health monitoring \& actuation everywhere (UC3), Living and working everywhere (UC4), Assistance from twinned cobots (UC5), and Sustainable food production (UC6)  \\ \hline
        Trust & Confidence in advanced digital systems during critical missions and in autonomous e-health decision-making & Emergency response \& warning systems (UC1) and Personal health monitoring \& actuation everywhere (UC3) \\ \hline
    \end{tabular}
    \label{tab:kvi_sns}
\end{table*}

\subsubsection{The COHERENT's approach to KVI definition}
The project aims to implement Proof of Concepts (PoCs) of Network Digital Twins (NDTs) ecosystems to provide future networks with management tools consistent with emerging global sustainability goals beyond simple energy sustainability. 
Furthermore, the project defines mechanisms to support the users of the networks in expressing their ``sustainable intents" through KVIs that will be mapped into KPIs and injected into the NDTs.
Following the approach of the COHERENT project, in which the authors of this paper participate to, starting from the variety of topics encountered in the analysis of the KVIs presented so far deriving from literature and design activities, Table \ref{tab:kvi_summary} provides an overview and a synthesis of those that could potentially be incorporated into PoCs aimed at demonstrating the feasibility in future Ethical networks of the various sustainable objectives (the Aims in the Table).
In the approach of COHERENT, the purpose is to define the KVIs as specific metrics to monitor the impact of KVs within an organization or a system.
They are intended to enable users to select Ethical network providers based on sustainability preferences, aligning the choices that the operator will have to make on its network infrastructures with user-driven sustainability goals.
Users can prioritize providers that demonstrate sustainable practices in line with their principles by declaring their level of sensibility toward Key Values.
However, it is crucial to ensure that KVIs remain a tool for authentic value-based decision-making and avoid turning them into a mechanism for creating unnecessary functionalities, which often serve profit-driven goals rather than real user needs.

\begin{table*}
\centering
\caption{Summary of the Proposed Main KVIs and Their Features in the Coherent Project}
\begin{tabular}{|P{1.8 cm}|P{5.8 cm}|P{4.8 cm}|P{4 cm}|}
\hline
\textbf{KVI} & \textbf{Aim} & \textbf{Metrics} & \textbf{References} \\ \hline
Fair Working Conditions & Promote labor policies to ensure the well-being of workers, regardless of religion, ethnicity, gender, origin, or any other personal condition & Workers' welfare and safety conditions, non-existence of child labor and slavery, presence of a gender equality plan & SDGs (T. 8.2, 8.3, 8.5, 8.7), ADROIT6G (D2) \\ \hline

Safe and Secure Working Environments & Companies should not have larger economic profits by using low-quality products that can put lives at risk or saving money by reducing safety or security measures & Presence of work safety measures, number of people affected by disasters due to insufficient safety and security measures & SDGs (T. 8.8) \\ \hline

Open Collaboration & Promoting research and knowledge sharing to balance global economic development and improve worldwide well-being
& Number and quality of partnerships established & ADROIT6G (E4) \\ \hline

Digital Inclusion & Making network access equitable for everybody, reducing the gap between those who have access and those who do not & Inclusion of diverse users in the design and management process, affordability, minimization of the digital divide & SDGs (T. 9.C), ADROIT6G (D3), Hexa-X (Digital Inclusiveness), 6G SNS IA (Cultural connection, Digital inclusion, Education) \\ \hline

Emission Reduction Percentage & Reduce greenhouse gas emissions and mitigate environmental impact over time, with a timeline of targets to benchmark progress & Ratio between CO\textsubscript{2} emissions of the year compared to a specific past period & SDGs (T. 7.1, 9.4, 12.c), ADROIT6G (E1), Hexa-X (Sustainability), 6G SNS IA (Environmental sustainability) \\ \hline

Energy Consumption Efficiency & Reduce overall energy consumption through process optimization and adoption of energy-efficient technologies and equipment & Energy consumption benchmarks and energy savings calculations relative to a baseline period & SDGs (T. 7.3), ADROIT6G (I5), Hexa-X (Sustainability), 6G SNS IA (Environmental sustainability) \\ \hline

Circular Economy & Measure the progress in adopting circular economy practices to optimize resource use and minimize waste generation & Waste diversion rates, circular design assessments, stakeholder engagement surveys & SDGs (T. 12.5), Hexa-X (Sustainability) \\ \hline

Trustworthiness & 
Building trust in a network considers certifications and public-opinion surveys & Robustness, security, transparency, explainability, and fairness & SDGs (T. 16.6), ADROIT6G (D4), Hexa-X (Trustworthiness), 6G SNS IA (Trust) \\ \hline
\end{tabular}
\label{tab:kvi_summary}
\end{table*}

\subsection{Embedding KVIs into an Ethical Network Architecture} 
\label{subsub:architecture}
The projects mentioned in Section \ref{sec:KVI_projects} also have common features in terms of proposed network architectures. In the following we highlight which are the basic elements that unite the architectures and which emerge as the most promising for the future deployment of Ethical network solutions that support sustainable and ethical Key Values.

\textit{Artificial Intelligence} (AI) and  \textit{Machine Learning} (ML) are key features across all three architectures, though implemented differently:
ADROIT6G focuses on a fully AI-driven distributed intelligence, HEXA-X applies AI/ML for spectrum management and network automation, and COHERENT uses Digital Twins that leverage Intelligent techniques to predict and optimize network behavior. The role that Artificial Intelligence and Machine Learning can play in future sixth-generation (6G) wireless systems appears crucial from the perspective of network management that responds to sustainability principles.
Areas related to sustainability in which AI/ML can be decisive are, for example, those improving energy efficiency, network optimization for safety in work environments and prevention of related risks, providing network resources in a fair way to the different customer profiles, and many others among those identified by the SDGs. 

The concept of \textit{Intent-Based Networking (IBN)} also emerges among the enabling technologies used in the context of the projects mentioned. 
An intent is knowledge that is fed to the system to learn the customer objectives and allow automatic decision-making and optimization.
Specifically, IBN can play a role in achieving the SDGs by simplifying the expression of user expectations in a declarative and natural form.
However, the conversion and standardization of intents to service and resource models is still an open challenge.

Furthermore, architectural solutions for future telecommunication systems increasingly tend to introduce flexibility through \textit{Network Virtualization} techniques and novel virtual network functions to customize network services according to the user's necessities \cite{Bernardos21}. This aspect can also play an important role in future Ethical network solutions since only virtualization techniques of network elements allow for an effective analysis of current network conditions and a rapid reconfiguration of the behaviors of the network itself in accordance with the different (and sometimes contrasting) KVI expressed by customers.
From this perspective, it is worth mentioning some emerging paradigms also used by some of the projects mentioned, such as Zero Touch Network \& Service Management and Network Digital Twins.

Specifically, the \textit{Network Digital} Twin (NDT) architecture is an advanced framework that creates a virtual replica of a physical network. A NDT employs advanced communication technologies to enable real-time information exchange between physical objects and their virtual twins, among virtual twins, and among physical objects. An NDT manages multiple DTs simultaneously to model a group of objects so that some intermediate processing results can be shared among the collaborative DTs. This cooperation approach greatly saves the processing time delay and energy consumption and helps to improve the modeling efficiency \cite{DTNAS}.
Enabling architectures of this concept has also been the object of intense standardization activities within the Internet Research Task Force (IETF) \cite{IETF-DTN23} and ITU \cite{ITU-DTN22}. In the sustainability context, the use of NDT allows for understanding in advance the impact of different service and network solutions. It also automates the management operations according to the desires of the various stakeholders. 

\textit{Zero-touch network \& Service Management} (ZSM) concepts can play an important role in ethical and sustainable networks as well. 
The ultimate goal of the ZSM concept is, in fact, to enable an autonomous network system capable of self-configuring, self-monitoring, self-healing, and self-optimizing based on policies and service level rules (which could also be expressed through sustainability-related intents) without human intervention \cite{LIYANAGE2022103362}.

\begin{figure*}
    \centering
    \includegraphics[width=1\linewidth]{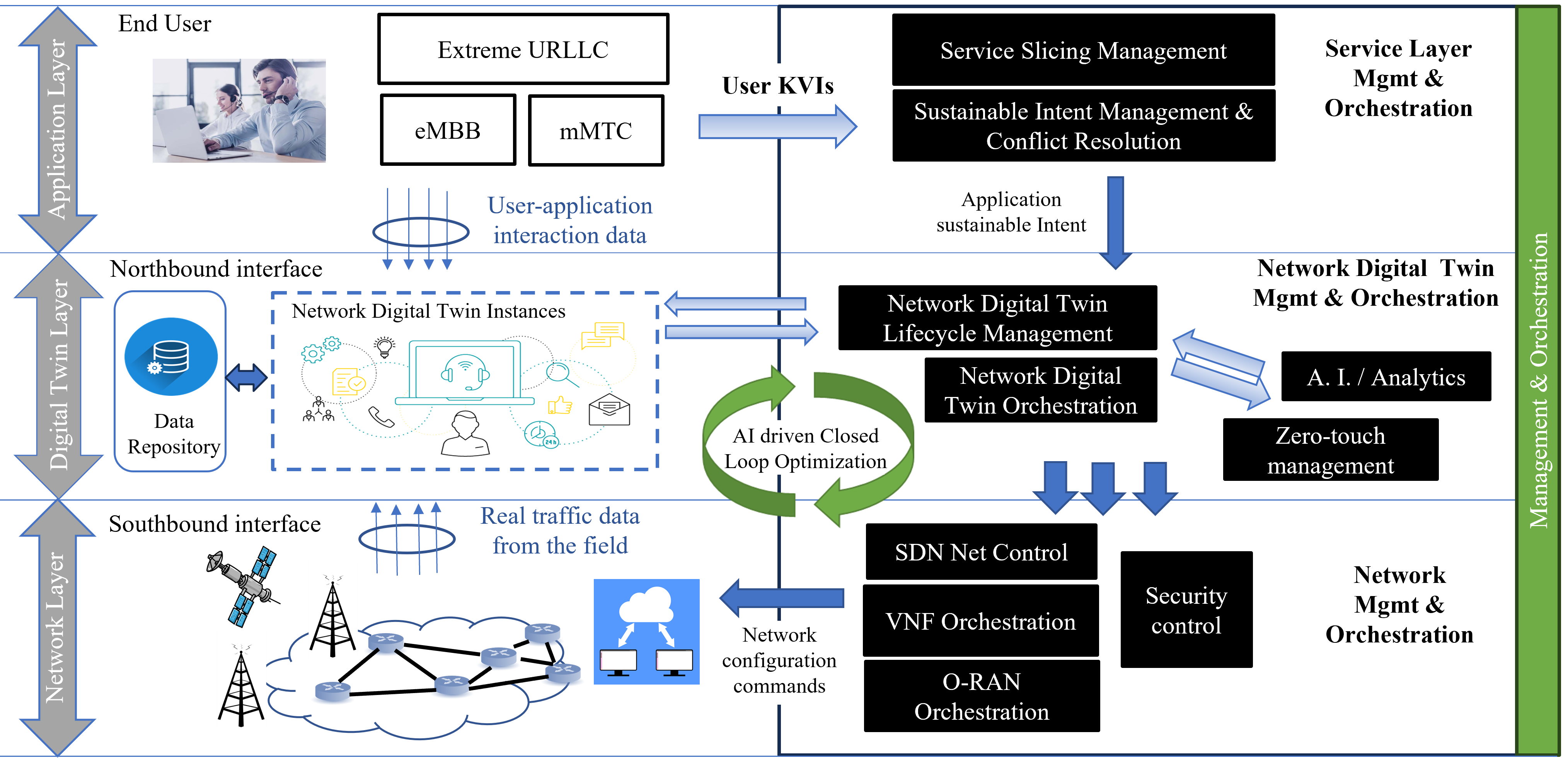}
    \caption{Network architecture that incorporates the main functionalities for the management of Key Values. Data-plane functions are shown on the left and management \& orchestration are shown on the right.}
    \label{fig:coherent_arch}
\end{figure*}

In light of what has been discussed, a very interesting potential convergence emerges between the paradigms of IBN, Network Virtualization, and DTN \cite{Verdecchia24} that represents a promising approach for the inclusion and management of KVI within future networks. 
This approach brings to a network architecture organized in three main layers, as depicted in Figure \ref{fig:coherent_arch}:
\begin{itemize} 
    \item The \textit{Application} Layer supports the user to express the intent via a human-friendly interface.
    Intents are profiled and refined by asking the user additional specifications, in case the initial intent is not sufficiently clear.
    \item The \textit{Digital-Twin} Layer provides the context to assist the modeling and translation of user intents. 
    Intents are then validated by the orchestrator that detects possible conflicts among requested services.
    \item The \textit{Network} Layer is the low-level infrastructure and physical resources of the network.
\end{itemize} 
The Application and the DT Layers are connected via a Northbound Interface that allows for intent injection and service exposure.
The DT and Network Layers use a Southbound interface to exchange data and control instructions.
To recap, users can define their goals through intents that are then associated with a model that profiles users, linking them with their Key Indicator preferences.
The orchestrator is responsible for verifying that the requests are in line with the user's profile, possibly asking for clarification if there are conflicts between the requirements.

\section{Discussion about Future Issues}
\label{sec:future_issues}
\begin{figure}
    \centering
    \includegraphics[width=1\linewidth]{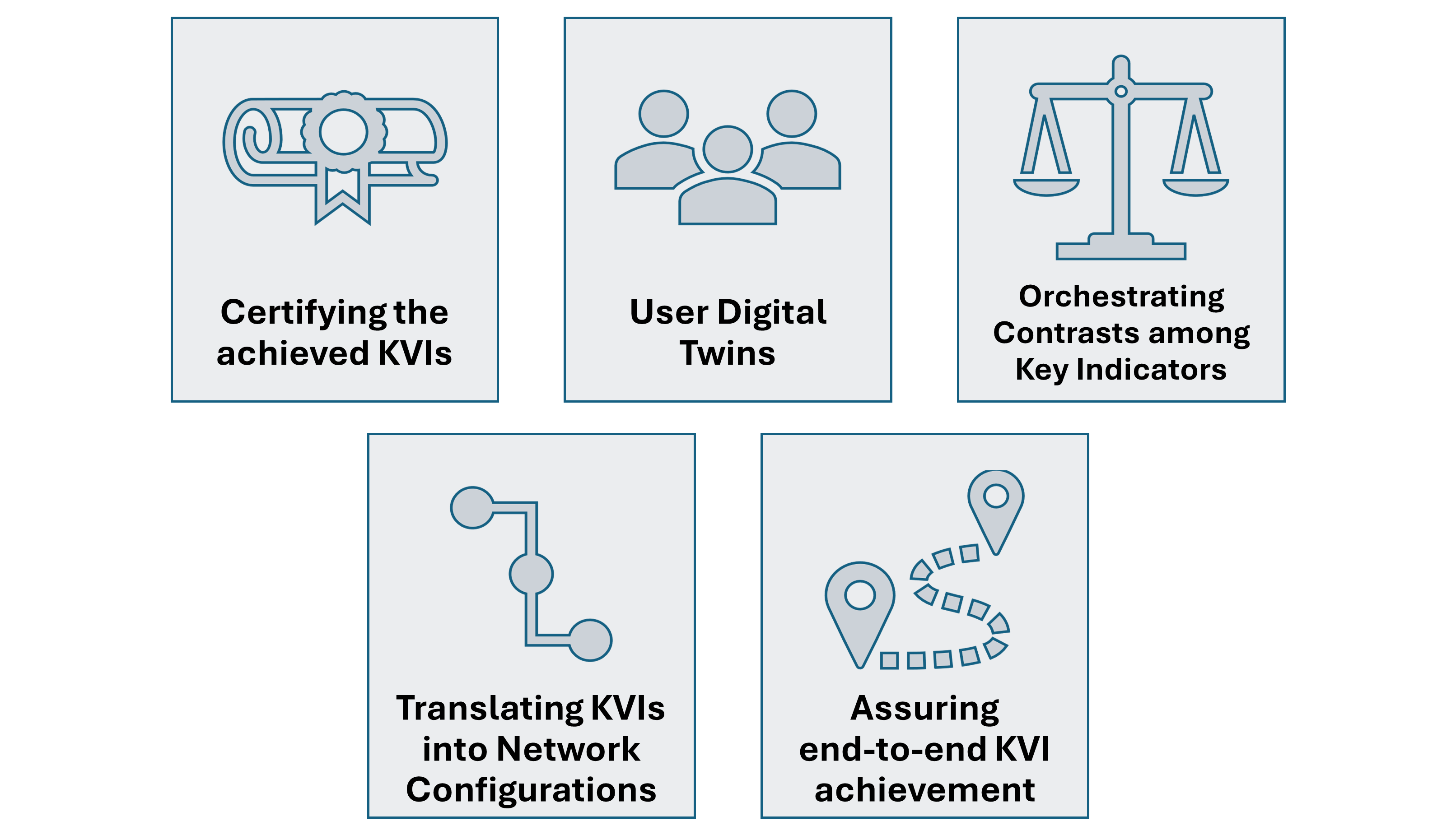}
    \caption{Future issues in Sustainable Networks}
    \label{fig:future_issues}
\end{figure}

There are five major issues that we believe need to be addressed in the short term period to make it possible to implement effective KVI-oriented network management. These are sketched in Fig. \ref{fig:future_issues} and discussed in the following subsections.

\subsection{Certifying the achieved KVIs}
We have discussed in Section \ref{subsec:users} the important role of the end-users in raising the attention of the companies in key values of their interest. Indeed, customers are becoming more and more interested in sustainability values while buying services from IT companies. 
Customers tend to request services that match their profile, so companies should leverage this opportunity to increase the service demand. Accordingly, the companies are expected to do their best to adopt these most demanded criteria while taking into account the economic implications. 
They need to analyze all the KVIs and their impact on customer opinion and reaction and then find a way to maximize their economic return. It means they should be able to measure and represent the achieved targets and communicate these properly to the users. 
This communication should be at least \textit{trustworthy} and \textit{clear} for a generic user so that 
the achievement of some of the key values could be provided by certification authorities that grant the implementation of the relevant certification if it exists (as discussed in Section \ref{sect:KVI_type}). 
Moreover, more than one certification can satisfy a single KVI.
For instance, SDG 7 (Affordable and Clean Energy) could be certified by different procedures, such as the ``ISO 50001, Energy management systems" and the ``ISO 52000 series of standards for the energy performance of buildings" \cite{Galluccio24}. 
However, not all of them can have a certification available, or the relevant certifications could be too generic to ensure the achievement of a given value in a precise way. 
This is the case of SDG 10 (Reduced Inequalities), in which the ``ISO Strategy 2016-2020" describes an action plan for developing countries but does not assure its achievement.
In all these cases, there is the need to define in a clear way how the achievement of the values of interest is measured and continuously monitored. 
Then, a proper, trustworthy procedure should be implemented so that the customers can rely on it.
Technologies exist, such as the usage of a central objective trust system of a distributed blockchain-based distributed ledger.

\subsection{User Digital Twins}
The service provider needs to know which are the sustainability interests of their customers by observing their behavior and service consumption. 
The user may also be interested as well to let the service provider know the preferences to be supplied with the best service options. 
This information would make it easier for the provider to inject into the service deployment procedures a view of their customer and find the optimal service configurations with a focus on sustainability. 
This result could be achieved by implementing User Digital Twins (UDTs), which are digital representations of the users that typically include information about their profile, preferences, service history, and status. 
By tracking and learning from the user, a UDT may help to define a list of KVI-based interests associated with the user and that may be fed into the IBN-based solution. 
Service orchestration should be driven by the DTN, where services, end users, and network segments (capable of hosting Virtual Network Functions or other services) are dynamically represented by their corresponding digital twins. These digital twins are expected to automatically interact with each other throughout the various phases of orchestration.
Clearly, this would be possible only if users allow the operator to collect and analyze their personal data, which, on the other hand, would benefit from easier interaction with the requested services.

\subsection{Orchestrating Contrasts among Key Indicators}
Conflicts can arise from variations in the goals of a single user or from differing interests between users and providers or among providers themselves. When a user's key value interest deviates from those of prior service requests, a compromise can be negotiated by asking for more details to adjust priority indicators or by identifying specific conditions where some indicators may be disregarded. 
In cases where interests of different entities (e.g., a customer and a provider, or two providers) clash, there is the need for a \textit{solver} that must decide which interest takes precedence or if different interests can be merged and a compromise can be found. 
For instance, a customer may want high throughput with low latency, while the provider might require higher fees or service limitations to meet this demand. Another example could involve balancing a service provider’s preference for renewable energy with the potential cost increase for selecting a certified energy provider.

In these scenarios, priorities can be determined through weights assigned to the need for achieving target values. A special case might involve recovery from environmental disasters, where urgent action is needed to minimize damage, even if it results in temporarily higher resource consumption. 
The literature emphasizes the importance of resolving conflicts between intents. Problem detection approaches \cite{Heorhiadi18}, including those using ML classifiers \cite{Jacobs19}, identify correlations or overlaps between interests by clustering users with similar application needs using K-means \cite{Bakhshi17}. 
Other conflict resolution methods include simple aggregation through intersections \cite{Comer18}, Policy Graph Abstraction (PGA) for merging multiple service graphs without conflicts \cite{Prakash15}, or classic optimization approaches \cite{Abhashkumar17}. One notable study focuses on a conflict detection and policy resolution algorithm that addresses three areas: connectivity, applications, and virtual machine creation, while also considering the temporal dynamics of intents \cite{Zheng22}.

\subsection{Translating KVIs into Network Configurations}
In an ethical-oriented network architecture, there is the need for a translator that converts the defined KVIs into detailed configurations and policies that can be executed by the underlying network and compute infrastructure. 
This translator should operate across different layers to ensure that high-level KVIs are effectively translated into actionable service configurations. 
Thus, translation involves decomposing KVIs into sub-indicators and delegating them to the appropriate components within the system. 
In this context, there are several issues. 
Firstly, the KVIs should be specified using standard encodings, including JSON, XML, and natural language processing techniques, that can easily be used to represent the requests of the customers or UDTs mentioned before. 
Whereas it could be possible that only a pre-defined list of KVI descriptors could be made available from which the desired KVIs should be selected. 
In some cases, it could be required to freely define the level of importance of the KVIs asked by the customers and select those that should be fulfilled. 
Secondly, the rules for the translations still need to be defined. There are no current studies that have provided a solution to map high-level sustainability descriptors in specific network configurations.

\subsection{Assuring end-to-end KVI Achievement}
Most of the time, a service is deployed across providers' and organizations' boundaries in a distributed and multi-tenant environment. 
Anyway, KVIs should be provided end-to-end, asking for collaboration among the involved parties. In current service provisioning scenarios, this is a procedure very often addressed so that specific agreements are set between collaborating provides. 
Whereas this collaboration can now rely on well-defined standardized metrics and methodologies, its implementation in terms of KVIs is completely new. 
A possible approach is to convert the end-to-end KVIs in terms of KPIs preliminarily and then conduct the collaboration in terms of already known rules. 
Still, in this case, there would be the need to distribute the contributions to the achievement of the desired KVIs (in terms of KPIs) among the involved operators, which would not be that straightforward.

\section{Conclusions}
\label{sec:conclusions}
Modern societies are increasingly aware of social and environmental challenges driven by crises related to climate change and social inequality. 
In response, governments and international organizations promote sustainable development initiatives, while consumers and investors play a crucial role in supporting companies with strong commitments to sustainability.
However, the market is diverse, and it is not possible to define common policies for all sectors. 
Some vertical markets, such as Energy Production and Consumption, are ahead of others in this respect and can be an inspiration for setting guidelines and standards for others.
Formalizing Key Values and Key Value Indicators to measure their achievement is essential to align user requirements for sustainability with service provider offerings.
With regard to telecommunications, this paper describes the approaches of three major flagship projects (ADROIT6G, HEXA-X, and COHERENT) for defining ethical network architectures that support KVIs.
This innovative research topic still has many open issues to address, including certification of KVIs, use of User Digital Twin to profile users, orchestration of contrasts between Key Indicators, translation of KVIs into network configurations, and assuring end-to-end KVI achievement.
Addressing these issues is critical to fostering progress in creating sustainable, ethically-driven telecommunications networks and could serve as a model for other industries.
Ultimately, ongoing research and collaboration will be essential for refining these frameworks and realizing their full potential in promoting sustainability across all sectors.

\section*{Acknowledgment}
This work was partially supported by the European Union - Next Generation EU under the Italian National Recovery and Resilience Plan (NRRP), Mission 4, Component 2, Investment 1.3, CUP F83C22001690001, CUP E63C22002070006, CUP C37G22000480001, partnership on “Telecommunications of the Future” (PE00000001 - program “RESTART”).

\section*{Declaration of generative AI and AI-assisted technologies in the writing process}
During the preparation of this work the authors used Grammarly in order to improve language and readability. After using this tool, the authors reviewed and edited the content as needed and took full responsibility for the content of the publication.

\bibliographystyle{IEEEtran} 
\bibliography{bibliography} 

\end{document}